\documentclass[utf8]{frontiersSCNS} 

\usepackage{url,lineno, microtype, subcaption}
\usepackage[onehalfspacing]{setspace}
\usepackage{xcolor}
\usepackage{caption}
\usepackage[finalnew]{trackchanges}
\usepackage{transparent}
\usepackage{colortbl}

\definecolor{bg}{rgb}{.96, .96, .96}

\def\keyFont{\fontsize{8}{11}\helveticabold }
\def\firstAuthorLast{Oliveira {et~al.}} %use et al only if is more than 1 author
\def\Authors{Denny M. Oliveira\,$^{1,2,*}$, 
             Eftyhia Zesta\,$^2$,
             Dibyendu Nandy\,$^{3,4}$}

\def\Address{$^{1}$Goddard Planetary Heliophysics Institute, University of Maryland, 
                   Baltimore County, Baltimore, MD, United States
             $^{2}$Geospace Physics Laboratory, NASA Goddard Space Flight Center, Greenbelt, MD, United States
             $^{3}$Department of Physical Sciences, Indian Institute of Science Education and Research Kolkata, Mohanpur, West Bengal, India
             $^4$Center of Excellence in Space Sciences India, Indian Institute of Science Education and Research Kolkata, Mohanpur, West Bengal, India
             }

\def\corrAuthor{Denny M. Oliveira}
\def\corrEmail{denny@umbc.edu}

\def\Bz{B$_\mathrm{z}$}

\usepackage[breaklinks, colorlinks = true,
            linkcolor = SkyBlue,
            urlcolor = SkyBlue,
            citecolor = SkyBlue,
            anchorcolor = white,
            breaklinks]{hyperref}

\begin{document}
\onecolumn
\firstpage{1}

%\title[Shock data base]{A Data Base of Interplanetary Shocks Observed by Wind and ACE at the Lagrangian Point L1} 
\title[Storm-time Starlink reentry]{The 10 October 2024 geomagnetic storm may have caused the premature reentry of a Starlink satellite} 

\author[\firstAuthorLast ]{\Authors} %This field will be automatically populated
\address{} %This field will be automatically populated
\correspondance{} %This field will be automatically populated

\extraAuth{}% If there are more than 1 corresponding author, comment this line and uncomment the next one.
%\extraAuth{corresponding Author2 \\ Laboratory X2, Institute X2, Department X2, Organization X2, Street X2, City X2 , State XX2 (only USA, Canada and Australia), Zip Code2, X2 Country X2, email2@uni2.edu}

\maketitle

\begin{abstract}

    In this short communication, we qualitatively analyze possible effects of the 10 October 2024 geomagnetic storm on accelerating the reentry of a Starlink satellite from \add{very} low-Earth orbit (\add{V}LEO). The storm took place near the maximum of solar cycle (SC) 25, which has shown to be more intense than SC24. Based on preliminary geomagnetic indices, the 10 October 2024, along with the 10 May 2024, were the most intense events since the well-known Halloween storms of October/November 2003. By looking at a preliminary version of the Dst index and \change{two-line element (TLE) altitude}{altitudes along with velocities extracted from two-line element (TLE)} data of the Starlink-1089 (SL-1089) satellite, we observe a possible connection between storm main phase onset and a sharp decay of SL-1089. The satellite was \change{scheduled}{predicted} to reenter on 22 October, but it reentered on 12 October, 10 days before schedule. The sharp altitude decay of SL-1089 revealed by TLE data coincides with the storm main phase onset. \add{We compare the deorbiting altitudes of another three satellites during different geomagnetic conditions and observe that the day difference between actual and predicted reentries increases for periods with higher geomagnetic activity.} Therefore, we call for future research to establish the eventual causal relationship between storm occurrence and satellite orbital decay. As predicted by previous works, SC25 is already producing extreme geomagnetic storms with unprecedented satellite orbital drag effects and consequences for current megaconstellations in VLEO.

    \tiny
    \keyFont{ \section{Geomagnetic storms, solar flares, thermospheric mass density, satellite orbital drag, satellite reentry}} 
\end{abstract}

\section{Introduction}

    Intense solar wind perturbations, such as coronal mass ejections (CMEs), can greatly disturb the Earth's magnetic field due to the occurrence of geomagnetic storms \citep{Akasofu1966,Gonzalez1994}. Geomagnetic storms are global phenomena characterized by intense magnetospheric energy input into the ionosphere-thermosphere (IT) system. \add{\protect{During active times, the primary energy sources that heat the atmosphere are Joule heating and particle precipitation \citep{Knipp2004,Prolss2011}.}} Such magnetospheric energy primarily \remove{enters the IT system} \change{leading}{leads} the thermosphere to heat and upwell globally \change{due to}{driving} the propagation of large-scale gravity waves and atmospheric wind surges \citep{Bruinsma2007,Prolss2011,Emmert2015}. \add{\protect{The atmosphere first responds to energy input at high latitudes within minutes of the storm main phase onset \citep{Shi2017}, whereas it responds globally $\sim$ 3 hours after storm main phase onset \citep{Oliveira2017c}}}. While flying in low-Earth orbit (LEO, below 1000 km altitude), due to increased levels of atmospheric density, satellites then experience enhanced levels of drag forces which in turn enhance orbital drag effects. Such effects are quantified by many parameters, including satellite geometry, drag coefficient, area-to-mass ratio, and thermospheric neutral mass density \add{\protect{and ions}}, which is derived from drag acceleration measurements \citep{Sutton2005,Chen2012,Mehta2023}. Thermosphere heating and subsequent orbital drag effects usually occur during geomagnetic storms of different levels, but they are much more intense during extreme events \citep{Krauss2015,Oliveira2019b,Zesta2019a}. \par

    SpaceX is a private company that has recently launched thousands of satellites into LEO. That satellite megaconstellation is named Starlink, with the primary goal to provide internet service worldwide \citep{Ren2022}. One of the most recent examples of storm-time satellite orbital drag effects experienced in LEO is provided by Starlink satellites. According to \cite{Hapgood2022}, 49 Starlink satellites were deployed on 3 February 2022 while a weak geomagnetic storm, classified as a G1 event, was \change{ranging}{raging} on. As a result, 38 satellites did not make it to their intended altitude due to storm effects. \change{At least by}{Prior to} early February 2022, SpaceX launched their satellites to \add{\protect{very-low Earth orbit (VLEO)}} altitudes near 210 km where electric thrusters were turned on to uplift the satellite to operational altitudes around 500 km \citep{Hapgood2022}. However, the satellites \change{used to}{typically} perform a few orbits before being lifted up, but the environment was quite perturbed due to the minor geomagnetic storm occurrence \citep{Dang2022,Fang2022,Berger2023}. \par

    If a weak geomagnetic storm can bring down satellites from \add{\protect{V}}LEO, what can an extreme event do? In this short communication, we briefly discuss possible effects of the extreme geomagnetic storm of 10 October 2024 on forcing the premature reentry of a Starlink satellite. Although our observations were performed with preliminary versions of and limited data sets, it is very likely that the storm cut short the reentry process of the satellite. However, solid causal relationships can only be achieved with further investigations using multi-data sets and conduction of numerical/empirical simulations. As predicted before, solar activity is increasing in the current solar cycle and they are already impacting satellite orbits in \add{\protect{V}}LEO, as recently shown by Starlink satellites.

 \add{\protect{

    \section{Data}

    We use OMNI IMF (interplanetary magnetic field) and selected solar wind parameter data, along with SYM-H geomagnetic index data, for the period 10-11 October 2024. Both OMNI and SYM-H data have resolution of 1 minute \citep{King2005,Iyemori1990}. \par

    Geomagnetic activity is also represented by long-term 1-hour Dst indices \citep{Sugiura1964a}. Although with different resolutions, both Dst and SYM-H are frequently used to expresses ring current activity and geomagnetic storm intensities \citep{Wanliss2006}. According to the \cite{WDC_Dst2015}, final versions of the Dst index are only available from 1957-2020, whereas a provisional version of the index is avaible from 2021-2023. The 2024 version of the index is termed real-time/quick-look index. The key difference between the provisional Dst index and the real-time Dst index is that the provisional index is a more accurate representation of geomagnetic activity because it undergoes additional quality checks and manual corrections for data errors, while the real-time Dst is a quicker, less refined measurement used for immediate monitoring and forecasting, potentially containing inaccuracies due to unverified raw data. For these reasons, \cite{WDC_Dst2015} recommends use of real-time (quick-look) Dst data only for forecasting, diagnostic, and monitoring purposes \url{(https://wdc.kugi.kyoto-u.ac.jp/dst_realtime/index.html)}. \par

    \begin{figure}
        \centering
        \includegraphics[width = 14cm]{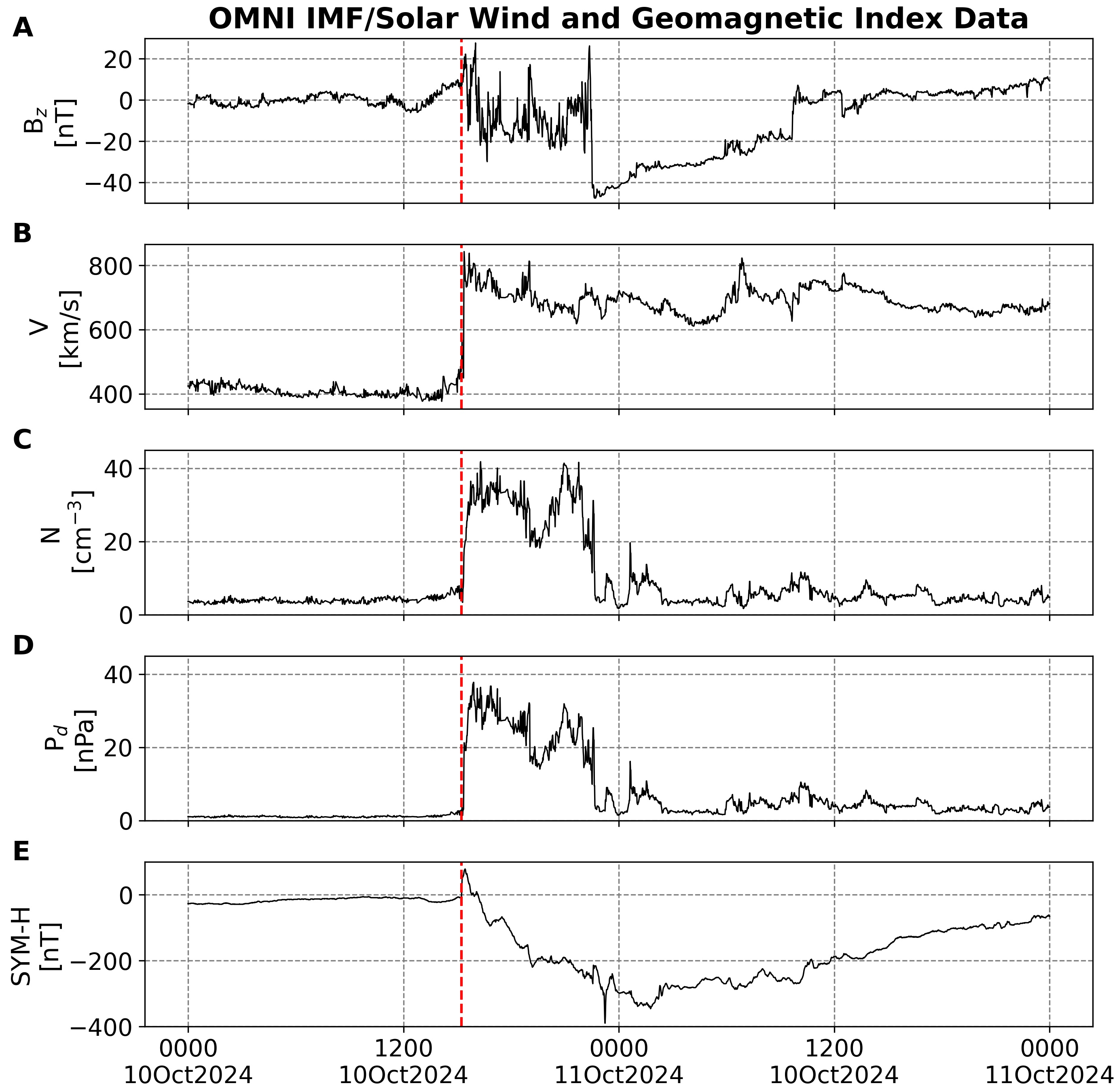}
        \caption{OMNI interplanetary magnetic field (IMF) and solar wind parameter profiles along with geomagnetic index data for the period 10-11 October 2024. ({\bf A}): IMF B$_\mathrm{z}$ component; ({\bf B}): solar wind flow speed; ({\bf C}): solar wind particle number density; ({\bf D}): solar wind ram pressure; and ({\bf E}): SYM-H geomagnetic index data. All data shown in the figure have resolution of 1 minute. The dashed red line indicates the storm sudden commencement onset on 10 October 2024 at 1514 UT.}
        \label{omni}
    \end{figure}

    For specification of satellites' orbital parameters represented by altitudes and velocities, we use Two-Line Element (TLE) data provided by the U.S. Space Force. TLE is a standardized format used to describe a satellite's orbital parameters, containing all the necessary information to calculate its position in space at a given time, presented in two lines of text. Each character in the file represents a specific orbital element like inclination, eccentricity, and mean anomaly, allowing for easy data exchange and prediction of a satellite's future path \citep[e.g.,][]{Kizner2005}. Altitudes and velocities are extracted from TLE data with the \texttt{pyephem} Python package (\url{https://pypi.org/project/pyephem/}). \texttt{pyephem} extracts satellite orbital parameters from TLE files based on physical principles of orbital mechanics, particularly Kepler's laws of planetary motion. The package then uses numerical methods to propagate the satellite's position over time, accounting for perturbative effects like gravitational anomalies, drag, and radiation pressure. This enables the accurate prediction of the satellite's trajectory and position at any given time based on the initial conditions provided in the TLE data \citep{Rhodes2010}. Satellite velocities are calculated from coverting the satellite's position from geodetic to cartesian coordinates, where $\vec{v} = d\vec{r} / dt$, with $\vec{dr} = (dx, dy, dz)$ being the geodetic displacement with respect to the Earth's center progated in a time step with $dt$ $\sim$ 10 seconds. The final altitude for each satellite is calculated using the corresponding decay epoch and TLE data. If a file for that particular decay epoch isn’t available, the latest TLE data file is used for the computations.
    }}

    % 1089: 2024-10-22 0:00:00
    % 2652: 2024-08-29 0:00:00
    % 1472: 2023-12-22 0:00:00
    % 2360: 2024-08-02 10:10:00

 \section{Observations}
    
    \remove{Figure 1, available at https://svs.gsfc.nasa.gov/14703/, shows a Solar Dynamic Observatory (SDO) image of a solar flare whose peak occurred at 0156 UT of 9 October 2024. The flare, originated from active region (AR) 3848, is indicated by the intense flash in the image at the center slightly above the Sun's equator. The figure shows a blend of light with wavelengths of 171, 304, and 131 \r{A}, which indicate extreme ultraviolet light. The solar flare is classified as an X1.8 flare, which belongs to the extreme edge of solar flare classifications (Bai and Sturrock; 1989; Schrijver and Siscoe, 2010; Saini et al., 2924)).} \par

    \begin{figure*}[]
        \centering
        \includegraphics[width = 14cm]{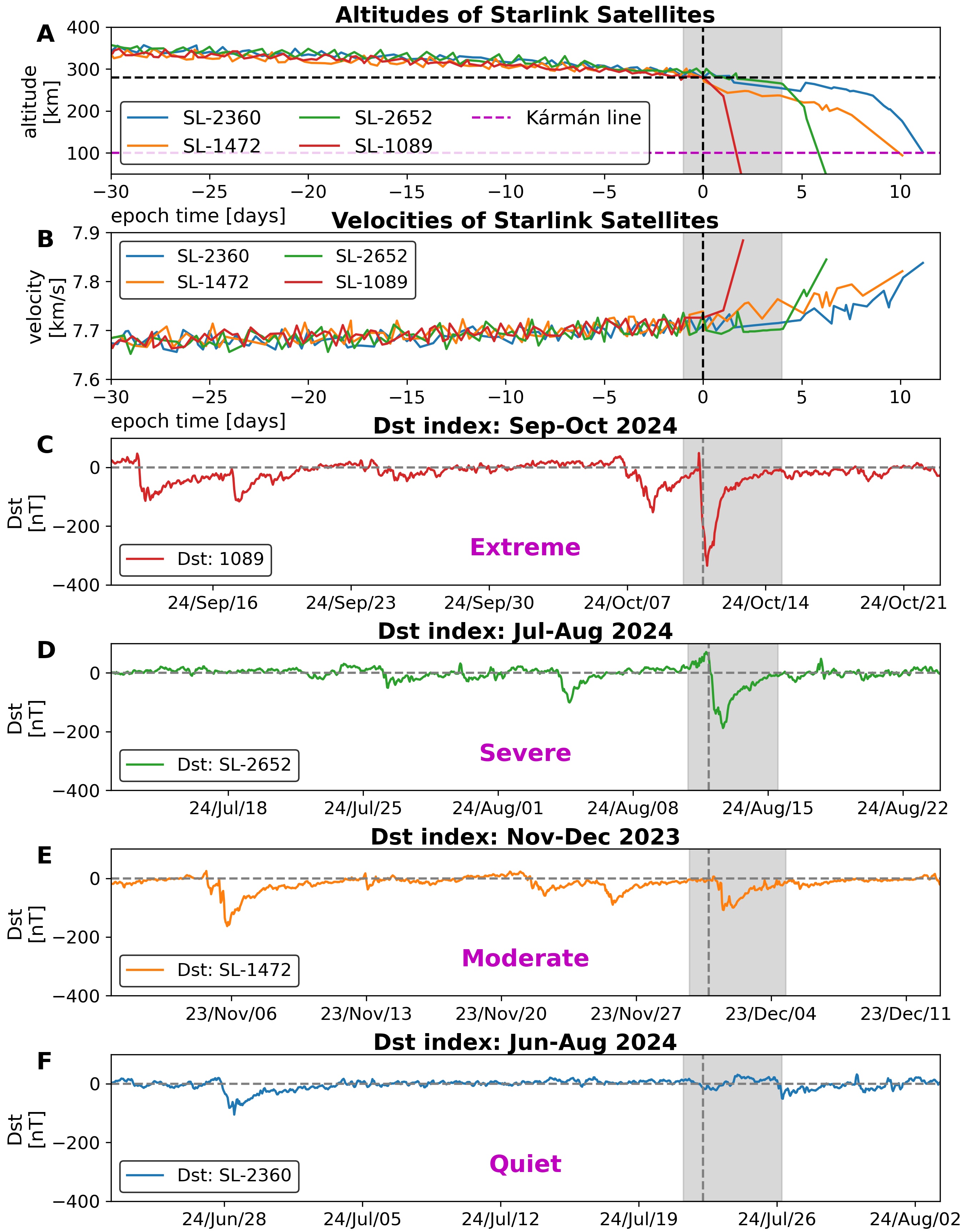}
        \caption{Altitudes/velocities and Dst data for four different Starlink satellites during their corresponding decay periods. ({\bf A}): Starlink altitudes; and ({\bf B}): Starlink velocities. The following panels indicate Dst index for the corresponding Starlink satellite. ({\bf C}): Sep-Oct 2024 (SL-1089); ({\bf D}): Jul-Aug 2024 (SL-2652); ({\bf E}): Nov-Dec 2023 (SL-1472); and ({\bf F}): Jun-Aug 2024 (SL-2360). The dashed vertical lines indicate a reference altitude at 280 km for each satellite. Data are plotted 30 and 10 days around the reference altitude. The grey highlighted area covers 1 to 4 hours around the reference altitude onset.}
        \label{starlink}
    \end{figure*}

    A CME associated with an X1.8 solar flare was observed by Solar and Heliospheric Observatory (SOHO) to be ejected from active region AR3848 on 9 October 2024 at 0212 UT (\url{https://kauai.ccmc.gsfc.nasa.gov/CMEscoreboard/prediction/detail/3670}). This particular CME impacted Earth on 10 October 2024 at around 1500 UT. Thus, the average speed of the CME on its way to 1 AU was $\sim$1200 km/s. Some of the space weather consequences of that CME are described below. \par

    % removed text
    \remove{
        Figure 2a shows the north-south component of the interplanetary magnetic field (IMF) vector observed by ACE (Advanced Composition Explorer) satellite (Smith et al., 1998)). The 5-minute averaged IMF B$_\mathrm{z}$ shows a positive jump from nearly null values to more than 20 nT on 10 October 2024 at around 1500 UT. Later, IMF B$_\mathrm{z}$ dips to values below --40 nT. The first (positive) B$_\mathrm{z}$ enhancement is due to the CME shock arrival, whereas the second (negative) enhancement is presumably due to the arrival of a magnetic cloud or magnetic material following the shock at the CME leading edge  (Illing and Hundhausen, 1985; Szabo et al., 2001; Zurbuchen and Richardson, 2006; Kilpua et al., 2019). Therefore, IMF observations by ACE indicate that the 10 October CME was an extremely intense solar wind driver  (Klein and Burlaga, 1982; Gonzalez and Tsurutani, 1987; Wang et al., 2003).
        } \par

    % added text
    \add{\protect{
        The IMF B$_\mathrm{z}$ component is shown in Figure\ref{omni}A. B$_\mathrm{z}$ stays relatively steady at nearly null values during 0000-1200 UT of 10 October, and ramps slowly up to values around 10 nT following a short period of negative values in the beginning of that day. As indicated by the red dashed vertical lines, the CME shock strikes the magnetosphere at 1514 UT of 10 October with a clear storm sudden commencement shown in Figure \ref{omni}E. A few minutes following the shock, \Bz{} jumps to values above 20 nT and later mostly shows negative values down to --20 nT. Nearly at the end of the storm main phase around 2225 UT of 10 October, \Bz{} plunges from $\sim$25 to $\sim$ --50 nT, which may have contributed to a long storm recovery of a few days. Figure \ref{omni}B indicates that the solar wind flow speed was mostly above 700 km/s. However, the solar wind number density (Figure\ref{omni}C) and dynamic pressure (Figure\ref{omni}D) stayed highly elevated (respectively above 30 cm$^{-3}$ and 20 nPa) only during most of the main phase, and showed relatively low values throughout the recovery phase. Figure \ref{omni}E indicates that the main phase ended at 11 October 0122 UT with minimum value of --335 nT, but it showed a --390 nT spike at 2314 UT of 10 October. Therefore, following the May 2024 extreme geomagnetic storm \citep{Hayakawa2024}, the October 2024storm was the second most intense extreme event since the October/November storms of 2023 \citep{Gopalswamy2005b}.
        }}

    % https://pypi.org/project/pyephem/

    % removed text
    \remove{
        Figure 2b shows quick-look (real-time) Dst data from 28 August 2024 to 15 October 2024. The hourly Dst data has been extensively used since its creation in 1957 during the International Geophysical Year (IGY) (Sugiura, 1964). The real-time (quick-look) Dst data is intended to be used for forecasting, diagnostic, and monitoring since its values are derived from provisional data. According to World Data Center for Geomagnetism, Kyoto et al. (2015), the real-time Dst data may contain noises and inaccurate baseline shifts that will be scrutinized later in the future before the release of the definitive Dst version (https://wdc.kugi.kyoto-u.ac.jp/dstrealtime/index.html). However, most likely the October 2024 storm will hold its status of an intense geomagnetic storm. 
    }

    % removed text
    \remove{
        It is clear from the figure that a classic storm sudden commencement (SSC) signature is shown by the quick-look Dst data. That event occurred at approximately 1500 UT of 10 October 2024. The SSC amplitude is 58 nT, and the minimum real-time Dst value is --341 nT, which occurred nearly 11 hours after SSC onset. This is compared to the other extreme storm of SC25: the event of May 2024. The SSC of that event occurred on 10 May at $\sim$1700 UT and had amplitude of 44 nT. The storm main phase developed in approximately 10 hours, with minimum quick-look Dst = --412 nT (Hayakawa et al., 2024). The most equatorward naked eye observations of the aurora reported by citizen scientists occurred in Chirimoyos (Mexico, N23$^\circ$26', W105$^\circ$48', 31.0$^\circ$ MLAT) in the northern hemisphere, and in the southern hemisphere, in Bromfield Swamp (Australia, S17$^\circ$22', E145$^\circ$33', --24.5$^\circ$ MLAT) (Hayakawa et al., 2024). 
    }

    \add{\protect{Unusually low-latitude auroras were seen during 10-11 October 2024.}} According to \url{aurorasaurus.org}, preliminary results show that the most equatorward observation of the aurora took place in the northern hemisphere near \change{Ochopee, Florida (N25$^\circ$54', W81$^\circ$19', 35.36$^\circ$ MLAT)}{Jerome, Florida (N25$^\circ$59', W81$^\circ$19', 35.36$^\circ$ MLAT)}, and in the southern hemisphere, near Fishers Hill, Australia (S32$^\circ$29', E151$^\circ$32', --41.74$^\circ$ MLAT)\footnote{This information was visually obtained by clicking on the date/time tabs on the top right of the screen. We then selected early hours of 11 October 2024 (e.g., 1:00am). The geographic coordinates and location names were then obtained from Google Maps and by zooming in and clicking on the nearest pinned location.}. Aurora observations at such low latitudes are quite common during extreme events \citep{Boteler2019,Hayakawa2019b,Bhaskar2020,Hayakawa2024}. More detailed and conclusive analyses of the most equatorward observations of the aurora during October 2024 using worldwide observations as performed \add{\protect{for the May 2024 event}} by \cite{Hayakawa2024} are under way. \par

    \begin{table}
            \centering
            \begin{tabular}{|l | c | c | c | c|}
                \hline
                                                & SL-1089           & SL-2652          & SL-1472            & SL-2360           \\
                \hline
                NORAD ID                        & 44967             & 48451            & 45736              & 47900             \\
                Launch epoch                    & 20/01/07 02:19    & 21/05/09 06:42   & 20/06/13 09:21     & 21/03/14 10:01     \\
                Last TLE file date              & 24/10/11 20:41    & 24/08/17 03:56   & 23/12/08 20:30     & 24/08/01 09:18    \\
                Epoch of reference altitude     & 24/10/10 20:00    & 24/08/11 22:00   & 23/11/30 18:11     & 24/07/22 06:00    \\
                \rowcolor{lightgray}
                Predicted reentry epoch         & 24/10/22 00:00    & 24/08/25 00:00   & 23/12/08 00:00     & 24/07/31 00:00    \\
                \rowcolor{lightgray}
                Reentry epoch                   & 24/10/12 20:41    & 24/08/18 03:56   & 23/12/10 20:30     & 24/08/02 09:18    \\
                Day difference in reentry epoch & --10              & --7              & --2                & +2                 \\
                Satellite age (years)           & 4.77              & 3.28             & 3.49               & 3.39              \\
                Quiet-time decay rate (km/day)  & 1.82              & 1.64             & 1.19               & 1.81               \\
                Storm-time decay rate (km/day)  & 128.52            & 38.05            & 18.34              & 15.78               \\
                Minimum Dst after $t_0$ (nT)    & --335             & --188            & --108              & --76               \\
                Geomagnetic activity level      & Extreme           & Severe           & Moderate           & Quiet                 \\
                \hline
            \end{tabular}
            \caption{IDs and orbital information of the four Starlink satellites used in this study. Satellite launch dates were obtained from \url{https://aerospace.org/reentries/} followed by the satellite NORAD ID. Estimates of the reentry dates are obtained from the last TLE by propagating predictions in time until the satellite reaches the K\'arm\'an line.}
            \label{table}
        \end{table}

        %                               Reentry                Predicted
        % 1089:                     2024-10-10 04:21    --> 2024-10-22 00:00        --> 10 days

        % 2652:                     2024-08-18 03:56    --> 2024-08-25 00:00        -->  7 days

        % 1472:                     2023-12-10 20:30    --> 2023-12-08 00:00        ---> 2 days
    
        % 2360: 2024-07-22 06:00    2024-08-02 09:18    --> 2024-07-31 00:00        ---> 2 days

    % added text
    \add{\protect{
        Altitudes and velocities for the satellites and Dst data are shown in Figure \ref{starlink}. The reference satellite is Starlink-1089 (SL-1089), whose deorbiting process was the most severe. We take a reference altitude as 280 km because SL-1089 started its sharp decay at that altitude during the storm. Another three satellites, SL-2652, SL-1472, and SL-2360, are chosen for comparisons due to three reasons: 1) the satellites reentered during different geomagnetic conditions; 2) they were around the reference altitude (horizontal dashed black line) near the onset of some geomagnetic activity (Figures \ref{starlink}C-F); and 3) the satellites had similar altitudes and orbital decay rates before reaching the reference altitude (Figure \ref{starlink}A; also see Supplementary Figures S1-4 for lifetime altitudes of all satellites). Thus, this suggests all four satellites were already reentering before reaching the reference altitude. The highlighted grey area corresponds to 1 day to 4 days around the reference altitude epoch time to represent quiet- and storm-times during the October 2024 event. In addition, as shown in Figure \ref{starlink}B, the velocity rate of each satellite generally increases as the satellite decays due to change of gravitational potential energy into kinetic energy. \par

        The satellites that passed through the reference altitude under different geomagnetic conditions are: SL-2360, quiet conditions; SL-1472, moderate conditions; SL-2652, severe conditions; and SL-1089, extreme conditions. Such storm classifications are based on minimum values of Dst during the storm period \citep{Zesta2019a}. Figure \ref{starlink}A shows that the more intense the geomagnetic activity level, the sharper the orbital decay rate. We consider a satellite renters the atmosphere when it crosses the K\'arm\'an line at $\sim$ 100 km (horizontal dashed magenta line), which is a commonly recognized boundary that defines the transition between Earth's atmosphere and outer space \citep{Karman1956,McDowell2018}. For each satellite, we define the predicted epoch, extracted from \url{space-track.org}, by searching each satellite under the tab ``Decay/Reentry" and by taking the reentry epoch predicted on the day the satellite crossed the reference altitude. As shown in Table \ref{table}, there is a difference between time delay, in days, determined as the day difference between the predicted and actual reentries of each satellite. The absolute day difference is larger for satellites reentering under high geomagnetic activity levels. During geomagnetic active times, the satellites reentered before prediction, except the satellite under quiet time conditions (SL-2360), which reentered 2 days after prediction. Table \ref{table} summarizes our findings and brings further information about all satellites.

    }}

        %\begin{table}
        %    \centering
        %    \begin{tabular}{c c c c c c c}
        %        \hline
        %            NORAD   & Sat  & Launch           & Last TLE           & Reentry              & Sat    &  Decay rate \\
        %            ID      & code & epoch            & file epoch         & epoch                & [years] & [km/day] \\
        %        \hline
        %        \rowcolor{lightgray}
        %            44967   & 1089 & 2020-01-07 20:57 & 2024-10-12 12:18:00 & 2024-10-12 20:41:03 & & \\
        %            46720   & 1823 & 2020-10-18 03:47 & 2022-08-31 08:13:00 & 2022-09-01 00:30:59 & & \\
        %        \rowcolor{lightgray}
        %            45736   & 1472 & 2020-06-12 09:00 & 2023-12-09 09:42:00 & 2023-12-10 19:37:45 & & \\
        %            47751   & 2168 & 2021-03-04 08:24 & 2024-08-04 00:39:00 & 2024-08-05 15:23:35 & & \\
        %        \hline
        %    \end{tabular}
        %    \caption{ID and orbital information of the five Starlink satellites used in this study. Satellite launch dates were obtained from \url{https://aerospace.org/reentries/} followed by the satellite code. Estimates of the reentry dates are obtained from the last TLE by propagating predictions in time until the satellite reaches the K\'arm\'an line.}
        %    \label{table} 
        %\end{table}

    % removed text
    \remove{
        Perhaps the most striking result of this paper is shown in Figure 2(c). The panel shows two-line element (TLE) altitude data of a specific satellite named Starlink-1089/2020-001BF, hereafter SL1089. SL1089 was launched into orbit on 11 October 2020 with orbital inclination of 53.04$^\circ$. Right after IMF B$_z$ assumes highly negative values and the storm main phase onset, SL1089 underwent a drastic and severe altitude decay. Although there were a couple of moderate storms in September 2024, drag effects caused by those events are not clearly seen in the figure. In addition, Figure 2(d) shows that the satellite's velocity starts to increase with orbital decay because its gravitational potential energy becomes kinetic energy (Prolss, 2011). Possible reasons and implications of this sudden orbital decay will be discussed below.
    }

    % https://www.starlink.com/updates

\section{Discussion, conclusion, and suggestions}
    
    Space Era is defined to have begun with the launching of Sputnik, the first satellite sent to space \citep{Launius2004}. Sputnik was launched in 1957, the solar maximum year of SC19 \citep{Clette2023}. Ephemeris data provided by one of the first Sputnik satellites was used by \cite{Jacchia1959} to arguably observe storm-time drag effects in LEO for the first time in history. Curiously, the highest yearly sunspot number observation also occurred in SC19 during 1957 \citep{Clette2023}. \cite{Zesta2019a} define geomagnetic storms with minimum Dst/SYM-H indices below the threshold of --250 nT to cause extreme effects on thermospheric neutral mass density and subsequent satellite orbital drag \citep{Oliveira2019b}, and it is well known that higher density enhancements result from more intense storms \citep{Krauss2018,Oliveira2019b,Krauss2020}. Although \cite{Meng2019} reported on the occurrence of nearly 40 extreme geomagnetic storms recorded with minimum Dst $<$ --250 nT since 1957, there are very few extreme events available to be studied with thermospheric neutral mass data derived from drag acceleration data collected by high-accuracy accelerometers. \cite{Oliveira2019b} and \cite{Zesta2019a} identified only 7 extreme events recorded by CHAMP and GRACE since early 2000's. As pointed out by \cite{Oliveira2021a}, this makes our understanding of extreme storm-time thermospheric dynamic response and subsequent enhanced satellite orbital drag effects quite limited. Therefore, more extreme geomagnetic storms are needed to enhance our understanding of the effects described above. \par

    \cite{Oliveira2021a} suggested two possible \change{paths}{approaches} that can be undertaken in order to improve our understanding of extreme satellite orbital drag: look into extreme events and superstorms of the past or expect for new extreme events in the upcoming solar cycles. The first approach was already taken by \cite{Oliveira2020b}, who studied drag effects in LEO on hypothetical satellites flying in the atmosphere during three historical superstorms (October/November 1903; September 1909; and May 1921) along with the well-known, space-era geomagnetic storm of March 1989. The authors defined two storm characteristics: storm intensity, defined by the minimum Dst value at the beginning of the recovery phase, and storm duration, defined as the time between the (presumably) impact of the driving CME on the magnetosphere and the end of the main phase. The authors concluded that storm duration can be more effective in comparison to storm intensity when determining the severity of drag effects, as occurred in the case of the March 1989 event. As pointed out by \cite{Bhowmik2018}, \cite{Nandy2021a}, \cite{McIntosh2023}, SC25 was expected to be stronger than SC24 based on sunspot number predictions. \change{This magnetic activity causally connects solar phenomena to direct space weather impacts around planetary objects such as the Earth and human technologies}{The magnetic activity causally connects solar phenomena to space weather effects around planetary objects such as the Earth and consequent impacts on human technologies} \citep{Nandy2023}. The National Oceanic and Atmospheric Administration (NOAA) announced \change{on 16 October 2024 that the Sun has reached the peak of SC25 on 15 October 2024}{in mid October 2024 that the Sun has reached the solar maximum period of SC25} \url{(https://tinyurl.com/3vcpt947)}. \remove{According to the Royal Observatory of Belgium, the maximum daily count of sunspot numbers in SC24 (167) occurred in January 2014, while the same for SC25 occurred on 16 October 2024 (https://www.sidc.be/SILSO/datafiles)} \change{As a result, two more extreme geomagnetic storms occurred}{So far, two extreme geomagnetic storms have occurred} in SC25: the event of May 2024 \citep{Hayakawa2024} and the event of October 2024. We expect more intense solar-driven geomagnetic storms to occur over the next few years around the peak of the current sunspot cycle. \par

    % SC24: 146

    The Starlink event of February 2022 taught us that even a minor/moderate geomagnetic storm can significantly enhance satellite orbital drag in \add{V}LEO. For instance, \cite{Fang2022} demonstrated with Starlink density data and empirical model results that the thermosphere was quite perturbed between the altitudes of 200 km and 400 km, with density increases of 50\%-125\% with respect to pre-storm values. Since the satellites were flying low in the thermosphere (at altitudes mostly around 200 km), the satellites encountered downfall before the thrusters were activated for further uplift to higher altitudes \citep{Hapgood2022}. As clearly seen in figure 1 of \cite{Oliveira2021a}, orbital drag effects dramatically increase at altitudes below 300 km (see CHAMP and GOCE altitudes during decommissioning) due to increasing density levels, even during quiet times. Therefore, decommissioning satellite operations deserve special attention for tracking during orbital decay, particularly during geomagnetic storms for safe and accurate reentry operations. \par
    
    SpaceX has made a commitment to safely de-orbit Starlink satellites in a time period of 2-5 years in order ``to keep[ing] space safe, sustainable and accessible, protect[ing] astronauts and satellites in orbit and the public on the ground", \add{as found in the document ``Commitment to Space Sustainability} (\url{https://api.starlink.com/public-files/Commitment%20to%20Space%20Sustainability.pdf}). Since SL-1089 was commissioned in January 2020, most likely it was already in reentry process at its fourth lifetime year when the October 2024 storm took place. As clearly seen in Figure \ref{starlink}A, SL-1089 decayed nearly 200 km within 48 hours (10 October to 12 October). Although the satellite was scheduled to be decommissioned on 22 October 2024, as clearly seen in the smooth altitude decay since early September 2024 and reported by \url{space-track.org}, the satellite reentered on 12 October 2024. \cite{Ashruf2024} attributed the losses of 12 Starlink satellites to space weather conditions surrounding the 10 May 2024 extreme geomagnetic storm. However, though those satellites decayed nearly 200 km in 3 days, the authors did not make it clear whether the satellites had already begun their reentries before the storm. As seen in Figure \ref{starlink}C, the extreme storm of October 2024 event had a minimum (quick-look) Dst value of --335 nT. Although the main phase of extreme geomagnetic storms tend to develop quite fast within a few hours \citep{Aguado2010,Cid2013}, the October 2024 event had a relatively long storm development duration ($\sim$11 hours). Such a combination of storm intensity and duration can cause extreme enhancements of thermospheric neutral mass density and subsequent orbital drag in LEO/VLEO \citep{Oliveira2020b}. As a result, the extreme geomagnetic storm of October 2024 presumably cut the reentry process of SL-1089 short by 10 days. This observation indicates that reentry operations of satellites should be monitored closely during storm times, particularly during extreme events. Such approach will help improve premature losses of satellites, accurate reentry locations, and effective collision-avoidance procedures. Such extreme orbital decay effects should be considered in the future since the number of extreme storms and the number of satellites in LEO/VLEO will be even larger in the years to come \citep{Oliveira2021a}. \par

    Finally, it should be pointed out that this is a preliminary analysis of the premature reentry of \change{Sarlink}{Starlink}-1089 in October 2024. This is due to the use of preliminary data \change{(quick-look Dst index, provisional ACE IMF data)}{(2024 quick-look/real-time Dst data)}, and the current lack of density data provided by the satellite. \add{Moreover, the use of less-refined Dst data is enough to determine the storm intensity (represented by minimum Dst values), and the occurrence time of storm main phase onset.} Therefore, we recommend further analyses of this event as performed before for the Starlink event of February 2022 by approaching data analyses and numerical/empirical model investigations \citep{Dang2022,Fang2022,Berger2023,Baruah2024}. \par

\section*{Data Availability Statement}

     Solar wind, IMF, and SYM-H data are provided by Space Physics Data Facility (SPDF) through the OMNIWeb Plus portal (\url{https://omniweb.gsfc.nasa.gov/}). Dst data, provided by \cite{WDC_Dst2015}, was obtained from \url{https://wdc.kugi.kyoto-u.ac.jp/dst_realtime/index.html}. The Starlink TLE data was downloaded from \url{space-track.org} by clicking on the tab ``ELSET Search". The website \url{space-track.org} is managed by the U.S. Air Force's 18th Space Defense Squadron (18 SDS), which is part of the U.S. Space Force. Login credentials are required to access the content provided by \url{space-track.org}.

\section*{Conflict of Interest Statement}

    The authors declare that the research was conducted in the absence of any commercial or financial relationships that could be construed as a potential conflict of interest.

\section*{Author Contributions}

    The TLE data analysis was performed by the first author. This perspective article was written by the first author. The manuscript was read and approved by the co-authors.

\section*{Funding}
    
    DMO acknowledges financial support provided by UMBC's START (Strategic Awards for Research Transitions) program (grant \# SR25OLIV). DMO and EZ thank the financial support provided by the NASA HGIO program through grant 80NSSC22K0756. DMO and EZ also acknowledge financial support by NASA's Space Weather Science Applications Operations 2 Research. DN's involvement in space weather research is sustained by the Ministry of Education, Government of India and a private philanthropic grant at the Center of Excellence in Space Sciences India, IISER Kolkata.

\bibliographystyle{frontiersinSCNS_ENG_HUMS} 
%\bibliography{/Users/dennyoliveira/Documents/Papers/Notes/Oliveira_main}

\end{document}